\documentclass[a4paper]{ifacconf}

\usepackage{graphicx,amsmath,url}      
\usepackage[round]{natbib}             

\def\portugues{0} 
\if\portugues0
   \usepackage[english]{babel}
  \else
   \usepackage[spanish,brazil,english]{babel}
\fi

\usepackage{multirow}
\usepackage{siunitx}
\usepackage{color}	
\usepackage{graphicx}	
\usepackage{amssymb}
\usepackage{amsmath,amssymb,amsfonts,textcomp}
\usepackage{mathtools}
\usepackage[table]{xcolor}
\usepackage{subcaption}
\usepackage{booktabs}

\usepackage{icomma}

\newenvironment{englishalgorithm}[1][]
  {\begin{algorithm}[#1]
     \selectlanguage{english}%
     \floatname{algorithm}{Algorithm}%
  }
  {\end{algorithm}}

\usepackage{multirow}

\usepackage[Algoritmo]{algorithm}
\usepackage[noend]{algpseudocode}
\usepackage{etoolbox}
\errorcontextlines\maxdimen

\makeatletter
\newcommand*{\algrule}[1][\algorithmicindent]{\makebox[#1][l]{\hspace*{.5em}\vrule height .75\baselineskip depth .25\baselineskip}}%

\newcount\ALG@printindent@tempcnta
\def\ALG@printindent{%
    \ifnum \theALG@nested>0
        \ifx\ALG@text\ALG@x@notext
            \addvspace{-3pt}
        \else
            \unskip
            \ALG@printindent@tempcnta=1
            \loop
                \algrule[\csname ALG@ind@\the\ALG@printindent@tempcnta\endcsname]%
                \advance \ALG@printindent@tempcnta 1
            \ifnum \ALG@printindent@tempcnta<\numexpr\theALG@nested+1\relax
            \repeat
        \fi
    \fi
    }%
\usepackage{etoolbox}
\patchcmd{\ALG@doentity}{\noindent\hskip\ALG@tlm}{\ALG@printindent}{}{\errmessage{failed to patch}}
\makeatother
\usepackage{tikz}
\usepackage{collcell}
 
\newcommand*{\MinNumber}{0.6449}%
\newcommand*{\MaxNumber}{1}%
 
\newcommand{\ApplyGradient}[1]{%
        \pgfmathsetmacro{\PercentColor}{100.0*(#1-\MinNumber)/(\MaxNumber-\MinNumber)}
        \hspace{-0.33em}\colorbox{gray!\PercentColor!black}{}
}
 
\newcolumntype{R}{>{\collectcell\ApplyGradient}c<{\endcollectcell}}
\renewcommand{\arraystretch}{0}
\usepackage[T1]{fontenc}

\usepackage[utf8]{inputenc}

\usepackage{ae}

\if\portugues1
 { }
 { }
 { }
 
\fi

\begin{document}

\if\portugues1

%
	
\begin{frontmatter}

\title{Integrating Time Series Classification into Railway Track Condition Assessment with Logistic-NARX Multinomial} 


\thanks[footnoteinfo]{Reconhecimento do suporte financeiro deve vir nesta nota de rodapé.}

\author[First]{Pedro H. O. Silva} 
\author[First]{Augusto S. Cerqueira} 
\author[Second]{Erivelton G. Nepomuceno}

\address[First]{Department of Electrical Engineering, Federal University of \\ Juiz de Fora (UFJF), Juiz de Fora, MG, Brazil, \\ (e-mail: silva.pedro@engenharia.ufjf.br, 
augusto.santiago@ufjf.edu.br).}
\address[Second]{Department of Electrical Engineering, Federal University of São João del-Rei (UFSJ), São João del-Rei, MG, Brazil, \\ (e-mail: nepomuceno@ufsj.edu.br).}

\selectlanguage{english}
\renewcommand{\abstractname}{{\bf Abstract:~}}
\begin{abstract}                
These instructions give you guidelines for preparing papers for the Sociedade Brasileira de Automática (SBA) technical meetings using the IFAC style. Please use this document as a template to prepare your manuscript in portuguese. For submission guidelines, follow instructions on paper submission system as well as the event website.

\vskip 1mm
\selectlanguage{brazil}
{\noindent \bf Resumo}:  As instruções abaixo são linhas gerais para a preparação de artigos para conferências e simpósios da Sociedade Brasileira de Automática (SBA) usando como base o estilo IFAC. Instruções de submissão podem ser encontradas no sistema de submissão de artigos ou no {\em website} do congresso.
\end{abstract}

\selectlanguage{english}

\begin{keyword}
Five to ten keywords separatety by semicolon. 

\vskip 1mm
\selectlanguage{brazil}
{\noindent\it Palavras-chaves:} Utilize de cinco a dez palavras-chaves separadas por ponto e vírgula.
\end{keyword}

\selectlanguage{brazil}

\end{frontmatter}
\else
%

\begin{frontmatter}

\title{Insightful Railway Track Evaluation: Leveraging NARX Feature Interpretation} 






\author[First]{Pedro H. O. Silva} 
\author[First]{Augusto S. Cerqueira} 
\author[Second]{Erivelton Nepomuceno}

\address[First]{Department of Electrical Engineering, Federal University of \\ Juiz de Fora (UFJF), Juiz de Fora, MG, Brazil, \\ (e-mail: silva.pedro@engenharia.ufjf.br, 
augusto.santiago@ufjf.edu.br).}
\address[Second]{Centre for Ocean Energy Research, Department of Electronic Engineering Maynooth University, Ireland, \\ (e-mail: erivelton.nepomuceno@mu.ie).}

\renewcommand{\abstractname}{{\bf Abstract:~}}   
   
\begin{abstract}                

The classification of time series is essential for extracting meaningful insights and aiding decision-making in engineering domains. Parametric modeling techniques like NARX are invaluable for comprehending intricate processes, such as environmental time series, owing to their easily interpretable and transparent structures. This article introduces a classification algorithm, Logistic-NARX Multinomial, which merges the NARX methodology with logistic regression. This approach not only produces interpretable models but also effectively tackles challenges associated with multiclass classification. Furthermore, this study introduces an innovative methodology tailored for the railway sector, offering a tool by employing NARX models to interpret the multitude of features derived from onboard sensors. This solution provides profound insights through feature importance analysis, enabling informed decision-making regarding safety and maintenance.

\end{abstract}

\begin{keyword}
machine learning; system identification; NARX models; time series classfication; wheel–rail contact dynamic forces; railroad dynamics.
\end{keyword}

\end{frontmatter}
\fi


\section{Introduction}

Time series analysis extends its significance into machine learning and classification tasks, where historical data patterns captured in time series empower algorithms to predict future events and classify data points effectively \citep{revin2023k}. In engineering applications, such as predictive maintenance in industrial machinery, time series data collected from sensors aids in forecasting equipment failures and optimizing maintenance schedules.


Among the challenges that need to be addressed in time series analysis, there is a necessity to establish methods that represent the most important system dynamics in a transparent and interpretable manner. Additionally, it is crucial to provide new approaches to data modeling problems by combining machine learning techniques with a simple and sparse structure. This is essential for many real-world applications where it is required that the resulting model be easy to interpret the system behavior while simultaneously possessing prediction performance.


An ideal candidate that can overcome such issues is the well-known NARMAX (Nonlinear Autoregressive Moving Average with Exogenous Inputs) model \citep{vidyala2024f, Aguirre}. This model can be constructed using input and output data from the system of interest, resulting in a final refined model that is typically simple and parsimonious. Additionally, the NARMAX model simultaneously addresses issues related to feature relevance, redundancy, and interaction \citep{billiings2019wei}.

In the past few years, there has been a surge in variations of the NARX methodology for modeling and prediction, aimed at improving the original algorithm's performance \citep{saadon2024f}. However, NARX models are primarily utilized in regression problems and binary classification tasks, where output variables consist of continuous or discrete-time sequences sampled from continuous processes \citep{AyalaSolares2017}. This limitation poses a challenge in tackling multiclass classification problems, where the output signal forms a multinomial sequence. Consequently, there is a notable opportunity to extend the application of NARX modeling to address more complex issues, such as real-world engineering problems \citep{Silva2020}.


One area that this work aims to delve into and contribute to is the railway sector, where data analysis is a particular challenge. This sector is facing increasing demands due to the need for more robust railway transportation services, resulting in heavier loads and faster trains. However, this increased intensity contributes to wear and tear on railroad infrastructure, highlighting the critical importance of early fault detection for safety and operational efficiency \citep{Silva2021b}. Contributing to track quality monitoring through dynamic measurements proves to be a valuable approach that provides essential information about the condition of the track \citep{sunlz2024f}. Integrating machine learning into the processing of this data to recognize patterns and anticipate emerging issues is essential for proactive and efficient maintenance. However, the conventional approach of direct analysis by specialized vehicles has its limitations in terms of speed, cost and impact on rail traffic.

Another crucial aspect is the necessity for trustworthy and Explainable AI (XAI) models to address the problem. In the railway sector, Explainable AI plays a pivotal role in interpreting the multitude of features derived from onboard sensors. By employing techniques such as feature importance analysis, analysts can gain profound insights into how various sensor-derived features influence system behavior. In an industry where safety, reliability, and efficiency are paramount, the ability to effectively interpret sensor data is critical for making informed decisions and devising proactive strategies  \citep{hassija2024}.

To address these limitations, this study proposes an approach that harnesses acceleration data and machine learning techniques based on NARX models to indirectly infer the state of the track \citep{marasco2024g}. Using data from multibody dynamics simulations, this research aims to develop predictive models capable of effectively assessing track conditions and providing a comprehensive understanding of the selected features and their importance in decision-making. This methodology allows for the indirect deduction of loads at the wheel–rail interface, incorporating safety limit information and creating a valuable tool for georeferencing decision-making in railway maintenance. 

\section{Nonlinear System Identification}\label{sec_ind}

System Identification is an experimental approach that aims to identify and adjust a mathematical model of a system, based on experimental data that record the behavior of system inputs and outputs \citep{Billings2013,Aguirre}. In particular, the interest in nonlinear system identification has received a lot of attention from researchers since the 1950s and many relevant results have been developed \citep{NM2016,Ferreira2017a}. A model representation constantly employed is the NARX model (\textit{Nonlinear AutoRegressive with eXogenous inputs}), consisting of a mathematical model based on differential equations.

\subsection{NARX Representation}

The NARX representation is a discrete-time model that explain the output value $y(k)$ as a function of previous values of the output and input signals:
\begin{equation}
\label{eqnarx}
\begin{aligned}
    y(k) = f^{l}(y(k-1),\cdots,y(k-n_y), \\
    u(k-1),\cdots,u(k-n_u))+e(k),
\end{aligned}
\end{equation}
\noindent where $f^l$ represents a nonlinear function of the model with nonlinearity degree $l \in \mathbb{N}$, $y(k)\in \mathbb{R}$ is the output of the system, and $u( k)\in \mathbb{R}$ is the input to the system in discrete time $k = 1, 2, \dots, N$; $N$ is the number of observations, $e(k)\in \mathbb{R}$ represents the uncertainties and possible noise in discrete time $k$, $n_y \in \mathbb{N}$ and $n_u \in \mathbb{N}$ describes the maximum lags for the output and input sequences, respectively. Most approaches assume that the function $f^l$ can be approximated by a linear combination of a predefined set of functions $\phi_i(\varphi(k))$, so that Equation (\ref{eqnarx}) can be expressed in the following parametric form:
\begin{equation}
\label{parnarx}
    y(k) = \sum_{i=1}^{m}\theta_{i}\phi_{i}(\varphi(k))+e(k),
\end{equation}
\noindent where $\theta_i$ are the coefficients to be estimated, $\phi_{i}(\varphi(k))$ are the functions that depend on the regression vector:
\begin{equation}
\label{vetregnarx}
\begin{aligned}
    \varphi(k)=[y(k-1),\cdots,y(k-n_y), \\ u(k-1),\cdots,u(k-n_u)]^\mathsf{T},
\end{aligned}
\end{equation}
\noindent where $\varphi(k)$ represents the previous outputs and inputs, and $m$ is the number of functions in the set. 
Finally, NARX models can be used to describe a wide variety of systems, simply obtaining analytical information about dynamic models. Another advantage is parsimony, meaning that a wide range of behaviors can be concisely represented using just a few terms from the vast search space formed by candidate regressors, as well as a small data set is needed to estimate a model, which can be crucial in applications where it is difficult to acquire a large amount of data.



\section{Logistic-NARX Multinomial Model Approach}


A hybrid multinomial classification method is presented, which allows the extraction and selection of features during the process. The method is based on the Orthogonal Forward Regression \citep{AyalaSolares2017} that selects the terms and combines the logistic function with the NARX representation to obtain a probability model:
\begin{equation}
    p(x) = \frac{1}{1+\exp{\left[ \sum_{m=1}^{M}\theta_{m}\phi_{m} \left(\varphi(k) \right)  \right]}}.
\end{equation}

Directly extending binary algorithms isn't always feasible, so many methods involve binarizing multiclass problems. One popular approach is One-Versus-All (OVA), which employs multiple binary classifiers to solve multiclass problems. Each classifier distinguishes one class from the rest, enabling classification based on a simple decision rule: 
\begin{equation}
\label{eq_multi}
    x \in w_v \Leftrightarrow \underset{1 \leq v \leq C }{\arg\max} \ f_v (x),
\end{equation}
\noindent defining $f_v$ as the result given by the model referring to the class $v$, meaning the probability between $0$ and $1$ of belonging to the v-th class with respect to an instance $x$.
\begin{englishalgorithm}
\small
\caption{Logistic-NARX Multinomial}
\begin{algorithmic}[1]
    \State \textbf{Input:}$\{y(k), k = 1, \dots, N \}$, $\mathcal{M} = \{\phi_i, i = 1, \dots, m\}$ , $l$, $n_y$, $n_u$, $k$ 
    \State \textbf{Output:} $\boldsymbol{\alpha} = \{\alpha_i, i = 1, \dots, k\}$, $\boldsymbol{\theta} = \{ \theta_i, i = 1,\dots,k\}$ 
    \For{$i=1:m$}
         \State $w_i \gets \frac{\phi_{i}}{\left \| \phi_{i} \right \|_{2}}$ 
        \State $r_{i} \gets $ Logistic regression accuracy in $w_i$ and $y$
         \EndFor
    \State $ j \gets \text{arg} \underset{1 \leq i \leq m }{\max}\{ r(w_{i},y) \}$ 
    \State $q_1 \gets w_j$
    \State $ \alpha_1 \gets \phi_{j}$
    \State Train logistic model with $\alpha_1$ and $y$
    \State Compute cross-validation
    \State Remove $\phi_j$ from $\mathcal{M}$
    \For{$s=2:k$}
         \For{$i=1:m$}
             \State $w_{i}^{(s)} \gets$ Orthogonalize $\phi_i$ in $[q_{_1},\dots,q_{_{(s-1)}}]$
             \If{$w_i^{\mathsf{T}}w_i<10^{-10}$}
             \State Remove $\phi_i$ from $\mathcal{M}$\;
             \State Next iteration\;
             \EndIf
             \State $r_{i} \gets $ Logistic regression accuracy in $w_i$ and $y$
         \EndFor 
         \State $ j \gets \underset{1 \leq i \leq m-s+1 }{\max}\{ r^{(i)}(w_{i},y) \}$ 
         \State  $q_s \gets w_j$
         \State  $ \alpha_s \gets \phi_{j}$
         \State Remove $\phi_j$ from $\mathcal{M}$
         \State $\boldsymbol{\alpha} \gets [\alpha_1,\dots,\alpha_{_{(s)}}]$
         \State Train logistic model with $\boldsymbol{\alpha}$ and $y$
         \State Compute cross-validation
    \EndFor 
    \State $\boldsymbol{\alpha} \gets [\alpha_1,\dots,\alpha_{_{(k)}}]$ \Comment{matrix of selected terms}
    \State $\boldsymbol{\theta} \gets [\theta_1,\dots,\theta_{_{(k)}}]$ \Comment{estimated coefficients vector}
\end{algorithmic}
\end{englishalgorithm}
Combining NARX models with multinomial classification involves adapting aspects of the Orthogonal Forward Regression (OFR) algorithm. Since OFR's error reduction rate metric isn't applicable to categorical outputs in multiclass settings, a logistic model using Maximum Likelihood Estimation (MLE) is employed instead. This model assesses the accuracy of predicting categorical variables from continuous ones, indicating their correlation. Accuracy is evaluated via K-Fold Cross Validation to measure model generalizability \citep{Silva2021b}.

In Algorithm $1$, line ($1$) represents the inputs composed by the vector $y(k)$ of classes, the matrix $\mathcal{M}$ constitutes the regressors formed by combinations of feature vectors, $k$ is the maximum number of selected terms and ($l$, $n_y$, $n_u$) are the parameters of the NARX model (Equation \ref{eqnarx}). Lines (3-8) aim to select the candidate terms $\phi_i$ with greater discriminatory power, based on the prediction accuracy of the logistic model. In lines (9-10) the logistic model is trained using the selected regressors $\alpha$ and the class vector $y(k)$, calculating the cross-validation of the classification. The selected terms are transformed at each step into a new group of orthogonal bases in lines (14-18), using the Gram-Schmidt Orthogonalization procedures. The process is repeated on lines (12-25) until it reaches the specified maximum number of model terms selection $k$. Finally, in lines (26-27) the matrix of selected terms $\boldsymbol{\alpha}$ and the vector of coefficients $\boldsymbol{\theta}$ of the estimated model are obtained. 

\section{Operational Risk Assessment on Railway Track}

\begin{figure}[ht!]
\begin{center}
\includegraphics[width=9cm]{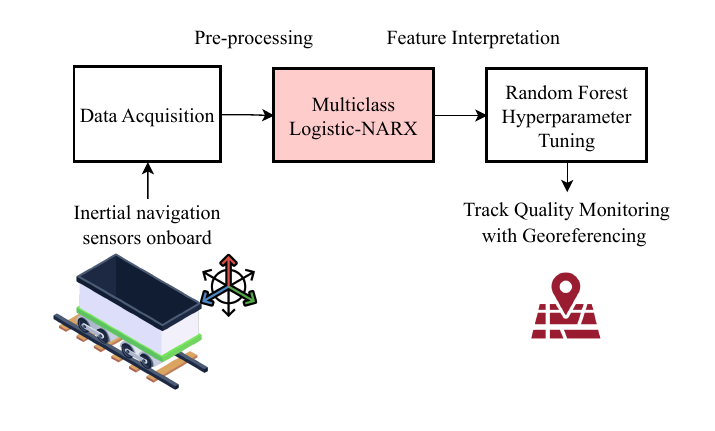}    
\caption{Workflow for railway track quality monitoring.} 
\label{fig_work}
\end{center}
\end{figure}

In this section, the workflow for track quality monitoring using inertial navigation sensors is presented. As shown in Figure \ref{fig_work}, the process starts with data acquisition by on-board sensors. The collected data is then pre-processed and fed into a Multiclass Logistic-NARX model. This is followed by feature interpretation by tuning the hyperparameters of a random forest algorithm. Finally, the results are used for georeferenced monitoring of track quality.

\subsection{Operational Safety Limits: Relationship of L/V Forces}

One of the main parameters used to ensure safety and prevent rail climbing derailment is the relationship between lateral and vertical forces at the wheel-rail contact. This relationship can be verified by the angle of the contact plane, which represents a limit between the force relationship at the contact. The limit is established by geometric relations of the forces projected on the contact plane, as represented by \cite{Nadal1908}:
\begin{equation}
    \label{eq_lv}
    \frac{\mathrm{L}}{\mathrm{V}} = \frac{\tan(\alpha) - \mu }{1+\mu \tan (\mu)}, \quad \quad \quad \quad  \mathrm{T} = \mu \mathrm{N},
\end{equation}
\noindent where $\mathrm{L}$ and $\mathrm{V}$ are the lateral and vertical forces, respectively, $\alpha$ is the angle of the contact plane, $\mu$ is the coefficient of friction between the contact parts, $\mathrm{T}$ is the tangential force, and $\mathrm{N}$ is the normal force. Equation \ref{eq_lv} is the most famous and widely used formulation in the railway field, serving as an indicator of operational safety through limits and critical values for the derailment coefficient.

Another mechanism that often occurs on straight tracks is derailment due to vertical force relief. In this scenario, there can be a reduction in the denominator $\mathrm{V}$ (vertical force) in the $\mathrm{L/V}$ ratio, leading to a increase in the active $\mathrm{L/V}$ indicator. Wheel relief is usually related to the movements of rigid bodies that cause changes in the distribution of vertical forces on the wagon. In scenarios where the wheel relief rate reaches high values, \textit{wheel lift} may occur, consisting of the wheel losing contact with the rail instantly.

The dependent variable of the proposed multiclass model aims to identify the critical conditions of the railway track according to the dynamics of wheel relief and $\mathrm{L/V}$. The classes were defined following a set of rules contained in the standards established by the FRA \citep{FRA2018}. The study adopts the standards to define safety limits, grouping the database by criticality in ascending order of severity: $\mathrm{Normal}$, $\mathrm{P2}$, $\mathrm{P1}$, $\mathrm{P0}$. The table \ref{tb_norms}, presents the limits determined by the standards based on wheel relief values, $\mathrm{L/V}$, and the conditions of loaded and empty wagons. 
\begin{table}[ht!]
\caption{Railway safety limits.}\label{tb_norms}
\begin{center}
\centering
\small
\setlength{\tabcolsep}{6pt} 
\renewcommand{\arraystretch}{1.2}
\begin{tabular}{l c c c c c c}
\hline
\multirow{2}{*}{\begin{tabular}[c]{@{}l@{}}Criticality\end{tabular}} & \multirow{2}{*}{\begin{tabular}[c]{@{}l@{}}Wheel relief\end{tabular}}  & \multicolumn{2}{c}{L/V}\\ \cline{3-4} 
 &  & loaded & empty\\ \hline
 $\mathrm{Normal}$ & $x \leq 50 \%$ & $x \leq 0.6  $ & $x \leq 0.6  $  \\\hline
$\mathrm{P2}$ & $ 50 \% \leq x < 60 \%$ & $-$ & $ 0.6  \leq x < 0.8 $  \\\hline
$\mathrm{P1}$ & $ 60 \% \leq x < 85 \%$ & $0.6  \leq x < 0.8 $ & $0.8  \leq x < 1.0 $ \\\hline
$\mathrm{P0}$ & $x \geq 85 \%$ & $0.8  \leq x < 1.0  $ & $x \geq 1.0 $  \\\hline
\end{tabular}
\end{center}
\end{table}
In summary, the safety limit parameters enable the labeling of criticalities, representing the four classes of the proposed problem. The goal is to use wheel relief rates and $\mathrm{L/V}$ to infer criticalities based on inertial sensor data and wagon operational information.

\subsection{Case Study: HPD Model}

\begin{figure}[ht!]
\begin{center}
\includegraphics[width=5.4cm]{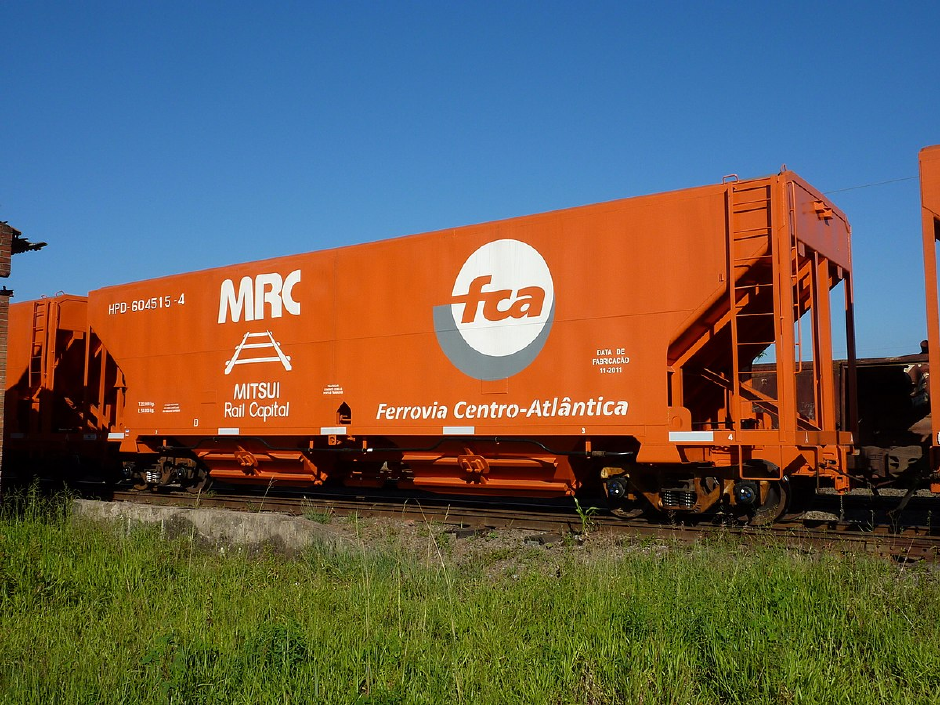}     
\caption{Railcar of the HPD (Hoppers) type.} 
\label{fig_sensor_3}
\end{center}
\end{figure}

In the case study, a wagon model of the HPD type (\textit{Hoppers}), which constitutes a significant portion of the assets of railway concessions worldwide, was considered. For the proposed problem, the HPD (\textit{Hopper car}) wagon (see Figure \ref{fig_sensor_3}) was modeled and instrumented for data acquisition in the study. The data sets were modeled in the VAMPIRE\textsuperscript{\textregistered} software (Vehicle Analysis Modeling Package in the Railway Environment), a commercial dynamic simulation (multibody) software that replicates the interaction between the railway vehicle and the railway track in the virtual environment, under certain boundary conditions based on real geometry, track stiffness, speeds, and the multibody model of the vehicle.

\begin{figure}[ht!]
\begin{center}
\includegraphics[width=5.5cm]{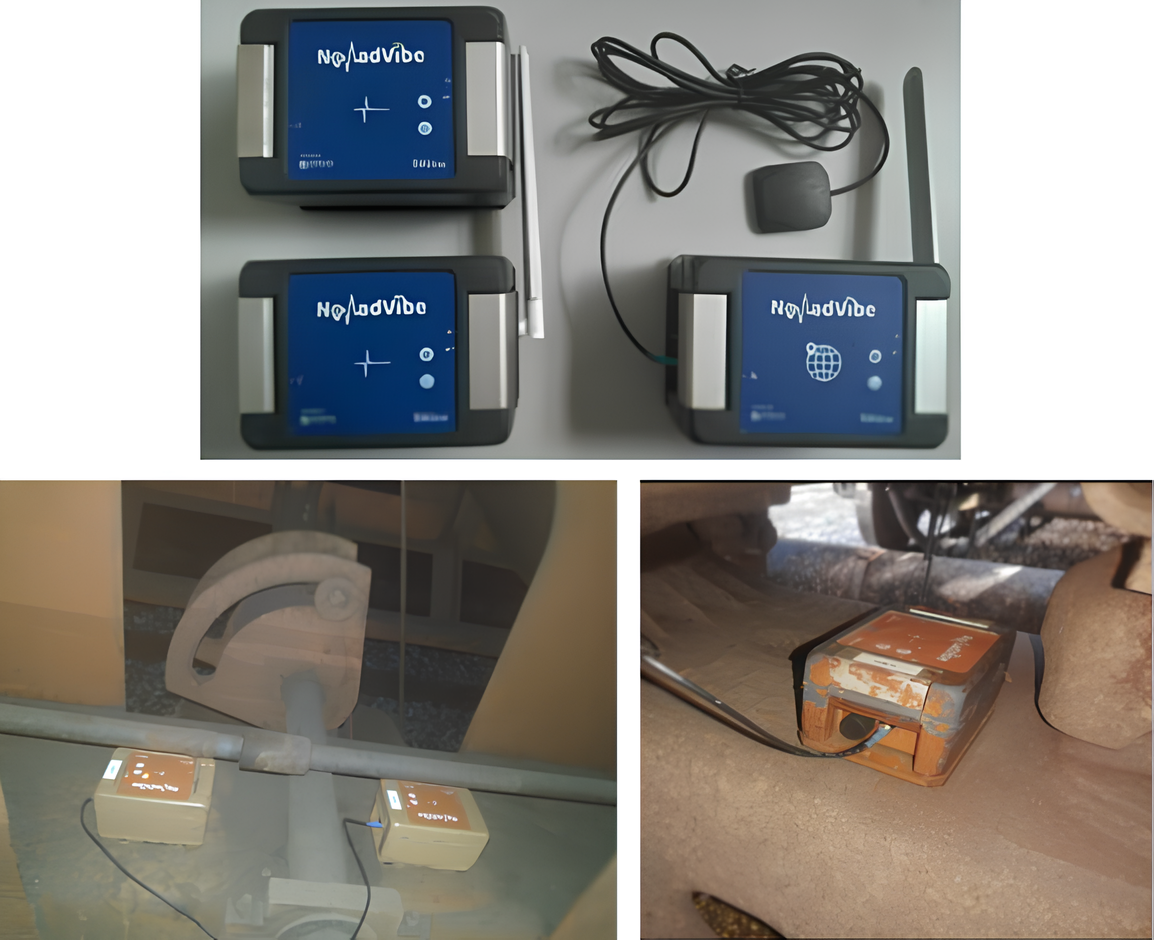}      
\caption{Equipment with inertial sensors.} 
\label{fig_sensor_2}
\end{center}
\end{figure}

As per the mentioned details, the railcar was also equipped with a set of accelerometers and gyroscopes, enabling, among other functions, the measurement of accelerations in the three axes ($X$, $Y$, $Z$) and angular velocities around them (Yaw, pitch, roll). The equipment also provides location information through GPS coordinates, speed, and the time of measurement, with a maximum acquisition frequency of $400$ $Hz$. The device, named NomadVibe, can be attached to metal surfaces using the magnets on its base, allowing for easy placement on the central crossmember of the truck and also on the wagon's side frame, as shown in Figure \ref{fig_sensor_2}.

\subsection{Data Acquisition and Pre-processing}

The input data for the model training are based on parameters and features collected by onboard inertial sensors and the VAMPIRE\textsuperscript{\textregistered} simulator, which relies on track geometry data to estimate dynamic wheel-rail contact responses. The collected and generated input data consist of wagon state parameters, speed scales, segments of the railway network, considering $10$ different sections with distinct operational characteristics and geometry to enrich the proposed model. For each combination of these mentioned parameters, features are extracted based on vertical and lateral acceleration and vibration modes, considering the car, leading truck, and trailing truck, as outlined.
The desired outputs, corresponding to the labels used in supervised learning model, consist of wheel relief rates and L/V information. In this case, the data includes wheel relief rates and L/V for the four wheels of the vehicles, with information estimated through the VAMPIRE\textsuperscript{\textregistered} simulator. During the study, a pivotal phase involved the execution of data cleaning, and standardization, meticulously addressing any irregularities present in the database \citep{FRA2018}. For further details, please refer to the work by \cite{silva2022cond}.

\subsection{Classifier for Track Condition Assessment}

A machine learning model or classifier is a computational system that learns to make predictions or decisions from data. It is trained using a process called supervised learning, where it uses a training dataset containing inputs (features) and their corresponding desired outputs (labels). The goal is to learn a mapping between inputs and labels so that the model can make accurate predictions on new unseen data.
The proposed methodology aims to use data from inertial sensors embedded in railway vehicles. These data are employed to determine dynamic responses, including acceleration and wheel-rail forces, concerning a specific excitation input. In the presented context, this excitation is caused by variations in the track geometry. Thus, the main objective of this work is to develop a methodology using machine learning that allows the inference of criticality levels associated with the railway track quality based on the obtained acceleration data.

\section{Results}

In this section, the NARX method will be employed to analyze data obtained from the assessment of railway track conditions, leveraging dynamic information gathered from railway vehicles. Additionally, we will explore its potential in addressing feature selection challenges encountered in classification problems.

The algorithms were executed on a machine equipped with an \textit{Intel Core i5}, CPU of $2.5 \ \text{GHz}$, and $16 \ \text{Gb}$ of RAM. All presented methods used the same datasets for training (\SI{80}{\percent}) and validation (\SI{20}{\percent}), employing $5$-fold cross-validation. For the proposed method, the following parameters will be utilized: a non-linearity degree of $l=2$ and a maximum number of selected terms of $n_{max}=10$. The algorithm selection will operate within a search space comprising $15$ terms. 


Table \ref{tb_model_rail} illustrates the score calculated during the term selection phase in the proposed algorithm, employing cross-validation. Consequently, the decision was made to select the model composed of $6$ terms to integrate the classifier. The details of the identified NARX model, denoted by $\hat{y}(k)$, including the terms, correlation values, and corresponding parameters, are comprehensively presented in the table.
\begin{table}[ht!]
\caption{Identified NARX model.}\label{tb_model_rail}
\begin{center}
\centering
\small
\setlength{\tabcolsep}{9pt} 
\renewcommand{\arraystretch}{1.2}
\begin{tabular}{lclr}
\hline
Model terms &  Score  & Feature & Parameter \\\hline
$u_{6}(k)u_{6}(k)$ &   $0.8068$  & \cellcolor{gray!15}Yaw\_box &  $0.6184$\\ \hline
$u_{5}(k)u_{5}(k)$ &   $0.4659$ & \cellcolor{gray!15}Acel\_vert\_t\_trail &  $-0.1357$\\ \hline
$u_{1}(k)u_{1}(k)$ &  $0.4164$ & \cellcolor{gray!15}Acel\_lat\_box & $-0,0164$\\ \hline
$u_{2}(k)u_{5}(k)$ &  $0.3919$ & \cellcolor{gray!15}Acel\_lat\_t\_trail & $0.0369$\\ \hline
$u_{4}(k)u_{13}(k)$ &  $0.3884$ & \cellcolor{gray!15}Acel\_vert\_box & $-0.0090$\\ \hline
$u_{2}(k)u_{3}(k)$ &  $0.3838$ & \cellcolor{gray!15}Type\_load & $0.0421$\\ \hline
\end{tabular}
\end{center}
\end{table}
The evaluation of the proposed method entails comparing it with other widely utilized techniques for handling classification problems. Table \ref{fig_conf_rail} depicts accuracy results in the form of intervals derived from cross-validation. Drawing from the confusion matrix presented in the table, the proposed method demonstrated a performance of \SI{74}{\percent} in classifying the (Normal class), while recording average results below \SI{60}{\percent} in the other classes. However, it is relevant to note that both SVM and KNN also showed similar results, with only the RF method capable of achieving results close to \SI{89}{\percent} for the (Normal class) and \SI{74}{\percent} for the (P0) class.
\begin{table}[ht!]
\small
\caption{Confusion matrix: methods.}\label{fig_conf_rail}
\centering
\hspace{-1.5cm}
\begin{subtable}[h]{0.15\textwidth}
\centering
\begin{center}
\setlength{\tabcolsep}{2pt} 
\renewcommand{\arraystretch}{1.3}
            \begin{tabular}{ccccc}
            \hline
                 &  $\mathrm{P0}$ & $\mathrm{P1}$ &  $\mathrm{P2}$  & $\mathrm{N}$\\ \hline 
                $\mathrm{P0}$  & \cellcolor{red!15}$61.6\%$  & $22.4\%$  & $11.9\%$ & $4.1\%$ \\ 
                $\mathrm{P1}$  &  $31.7\%$ & \cellcolor{red!15}$49.4\%$ &  $15.2\%$ & $3.7\%$ \\ 
                $\mathrm{P2}$ & $20.2\%$  &  $12.2\%$ & \cellcolor{red!15}$60\%$ & $7.6\%$ \\ 
                $\mathrm{N}$ & $8.6\%$  &  $8.3\%$ & $8.2\%$ & \cellcolor{red!15}$74.9\%$ \\
              \hline
            \end{tabular}
        \end{center}
        \caption{Logist-NARX.}
        \end{subtable}
         \hspace{1.5cm}
\begin{subtable}[h]{0.15\textwidth}
\centering
\begin{center}
\setlength{\tabcolsep}{2pt} 
\renewcommand{\arraystretch}{1.3}
            \begin{tabular}{ccccc}
            \hline
                 &  $\mathrm{P0}$ & $\mathrm{P1}$ &  $\mathrm{P2}$  & $\mathrm{N}$\\ \hline 
                $\mathrm{P0}$  & \cellcolor{red!15}$73.2\%$  & $16.2\%$  & $7.8\%$ & $2.8\%$ \\ 
                $\mathrm{P1}$  &  $13.6\%$ & \cellcolor{red!15}$66.7\%$ &  $15.8\%$ & $3.9\%$ \\ 
                $\mathrm{P2}$ & $8.1\%$  &  $14.4\%$ & \cellcolor{red!15}$69.6\%$ & $8\%$ \\ 
                $\mathrm{N}$ & $1.8\%$  &  $4.5\%$ & $4.1\%$ & \cellcolor{red!15}$89.6\%$ \\
              \hline
            \end{tabular}
        \end{center}
        \caption{Random Forests.}
        \end{subtable}
        
\begin{subtable}[h]{0.15\textwidth}
\centering
\begin{center}
\setlength{\tabcolsep}{2pt} 
\renewcommand{\arraystretch}{1.3}
            \begin{tabular}{ccccc}
            \hline
                 &  $\mathrm{P0}$ & $\mathrm{P1}$ &  $\mathrm{P2}$  & $\mathrm{N}$\\ \hline 
                $\mathrm{P0}$  & \cellcolor{red!15}$65.6\%$  & $26\%$  & $7.1\%$ & $1.3\%$ \\ 
                $\mathrm{P1}$  &  $45.3\%$ & \cellcolor{red!15}$30.9\%$ &  $19\%$ & $4.9\%$ \\ 
                $\mathrm{P2}$ & $31\%$  &  $30.6\%$ & \cellcolor{red!15}$34.4\%$ & $4\%$ \\ 
                $\mathrm{N}$ & $6\%$  &  $12\%$ & $7.6\%$ & \cellcolor{red!15}$74.4\%$ \\
              \hline
            \end{tabular}
        \end{center}
        \caption{SVM.}
        \end{subtable}
         \hspace{1.5cm}
\begin{subtable}[h]{0.15\textwidth}
\centering
\begin{center}
\setlength{\tabcolsep}{2pt} 
\renewcommand{\arraystretch}{1.3}
            \begin{tabular}{ccccc}
            \hline
                 &  $\mathrm{P0}$ & $\mathrm{P1}$ &  $\mathrm{P2}$  & $\mathrm{N}$\\ \hline 
                $\mathrm{P0}$  & \cellcolor{red!15}$69.2\%$  & $16.1\%$  & $10.4\%$ & $4.3\%$ \\ 
                $\mathrm{P1}$  &  $21.3\%$ & \cellcolor{red!15}$57.7\%$ &  $15.4\%$ & $5.5\%$ \\ 
                $\mathrm{P2}$ & $17.7\%$  &  $13.7\%$ & \cellcolor{red!15}$58.5\%$ & $10.1\%$ \\ 
                $\mathrm{N}$ & $6\%$  &  $11.1\%$ & $8.7\%$ & \cellcolor{red!15}$74.2\%$ \\
              \hline
            \end{tabular}
        \end{center}
        \caption{KNN.}
        \end{subtable}
        \hspace{1.3cm}
    \end{table}
In scrutinizing the findings presented in Table \ref{tb_desp_rail}, it is discernible that the proposed methodology achieved similar results to the other methods, excluding the RF method, which proved to be more suitable for the characteristics of the problem. However, the proposed method significantly improves interpretability by offering insights through coefficients, elucidating the impact of the features used in training on the prediction of classes (see Table \ref{tb_model_rail}). This feature contributes to a more thorough comprehension of the model's applicability in engineering contexts.
\begin{table}[ht!]
\caption{Performance comparison.}\label{tb_desp_rail}
\begin{center}
\small
\setlength{\tabcolsep}{6.5pt} 
\renewcommand{\arraystretch}{1.2}
\begin{tabular}{lccccc}
\hline
 & L-NARX M & RF & SVM & KNN  \\\hline
Average Accuracy & \cellcolor{gray!15}0.6073 &  0.7458 & 0.4657 & 0.6446 \\ \hline
Sensitivity & \cellcolor{gray!15}0.6148 & 0.7476 & 0.5131 & 0.6490 \\ \hline
Specificity & \cellcolor{gray!15}0.8743 & 0.9152 & 0.8438 & 0.8825 \\ \hline
Precision & \cellcolor{gray!15}0.6073 & 0.7458 & 0.4657 & 0.6446 \\ \hline
F1 Score & \cellcolor{gray!15}0.5864 & 0.7462 & 0.3754 & 0.6423 \\ \hline
\end{tabular}
\end{center}
\end{table}
Although the proposed method did not achieve superior accuracy results, the obtained model can be utilized as a feature extractor, resulting in a dimensionality-reduced dataset. The reduced dataset consists of $6$ features, selected from the initial set of $18$ features added to the problem. This dataset can facilitate gaining a better understanding of results and contribute to enhanced accuracy by mitigating overfitting to the data sample. Thus, by utilizing the selected features, we can employ the Random Forest model, optimizing its performance through hyperparameter tuning, to achieve better results for the given problem.

\subsection{Random Forest with Hyperparameter Tuning}

The selected model, Random Forest, was implemented through hyperparameter tuning and optimization. Hyperparameter tuning aims to find an optimal set of parameters that minimizes a predefined error function on the provided independent data. Additionally, cross-validation was employed to estimate the model's generalization performance.

Table \ref{fig_conf_pre_sens} presents the sensitivity matrix observed during the model validation, emphasizing the correctly identified positive cases. A detailed analysis of the matrix indicates that sensitivity in the $\mathrm{P0}$ criticality class exhibited the lowest values, possibly attributed to the limited availability of samples for training within this class. Regarding the $\mathrm{Normal}$ class, prediction errors are distributed between the $\mathrm{P2}$ and $\mathrm{P1}$ classes, with a minimal prediction error observed for the $\mathrm{P0}$ class in positive cases of $\mathrm{Normal}$. Conversely, the remaining classes exhibited robust sensitivity results, surpassing \SI{90}{\percent}.

In performance evaluation, accuracy is not a robust indication of the model's performance, given the significant class imbalance in the dataset. The most relevant metrics for model validation are sensitivity and precision. Sensitivity is crucial in situations where False Negatives (FN) are considered more detrimental than False Positives (FP). In the proposed model, prioritizing the detection of track irregularities, even if it results in classifying some normal conditions as critical (False Positive situation), is essential. Therefore, the model should exhibit high sensitivity, as misclassifying critical conditions as normal could pose a risk to the operational safety of railway vehicles.
\begin{table}[ht!]
\small
\caption{Matrix: sensitivity and precision.}\label{fig_conf_pre_sens}
\centering
\begin{subtable}[h]{0.15\textwidth}
\centering
\begin{center}
\setlength{\tabcolsep}{3pt} 
\renewcommand{\arraystretch}{1.3}
            \begin{tabular}{ccccc}
            \hline
                 &  $\mathrm{P0}$ & $\mathrm{P1}$ &  $\mathrm{P2}$  & $\mathrm{N}$\\ \hline 
                $\mathrm{P0}$  & \cellcolor{red!15}$74\%$  & $8.6\%$  & $4.6\%$ & $13\%$ \\ 
                $\mathrm{P1}$  &  $0.36\%$ & \cellcolor{red!15}$91\%$ &  $5.3\%$ & $3.6\%$ \\ 
                $\mathrm{P2}$ & $0.12\%$  &  $2.5\%$ & \cellcolor{red!15}$93\%$ & $4.2\%$ \\ 
                $\mathrm{N}$ & $-$  &  $0.57\%$ & $4.2\%$ & \cellcolor{red!15}$95\%$ \\
              \hline
            \end{tabular}
        \end{center}
        \caption{Sensitivity.}
        \end{subtable}
         \hspace{1.5cm}
\begin{subtable}[h]{0.15\textwidth}
\centering
\begin{center}
\setlength{\tabcolsep}{3pt} 
\renewcommand{\arraystretch}{1.3}
            \begin{tabular}{ccccc}
            \hline
                 &  $\mathrm{P0}$ & $\mathrm{P1}$ &  $\mathrm{P2}$  & $\mathrm{N}$\\ \hline 
                $\mathrm{P0}$  & \cellcolor{red!15}$93\%$  & $1.4\%$  & $0.2\%$ & $-$ \\ 
                $\mathrm{P1}$  &  $1.9\%$ & \cellcolor{red!15}$61\%$ &  $1\%$ & $-$ \\ 
                $\mathrm{P2}$ & $0.82\%$  &  $2.3\%$ & \cellcolor{red!15}$25\%$ & $-$ \\ 
                $\mathrm{N}$ & $4.5\%$  &  $35\%$ & $74\%$ & \cellcolor{red!15}$99\%$ \\
              \hline
            \end{tabular}
        \end{center}
        \caption{Precision.}
        \end{subtable}
         \hspace{1.6cm}

    \end{table}
    
Table \ref{tb_data} furnishes a comprehensive summary of the metric results derived from the model validation. As per the presented outcomes, the harmonic mean between precision and sensitivity (F1-Score) ranges from $0.73$ to $0.98$, with the exception of the $\mathrm{P2}$ class. This discrepancy suggests that the criticality class $\mathrm{P2}$ may not be well-defined in the problem, warranting reconsideration for a more realistic grading of criticalities.
\begin{table}[ht!]
\caption[General report of performance metrics.]{General report of performance metrics in the validation of different classes.}\label{tb_data}
\begin{center}
\small
\setlength{\tabcolsep}{10pt} 
\renewcommand{\arraystretch}{1.2}
\begin{tabular}{lccc}
\bottomrule
 & Precision & Sensitivity & F1-Score  \\\bottomrule
$\mathrm{P0}$ & \cellcolor{gray!15} 0.9281	 &  0.7428 & \cellcolor{gray!15} 0.8251 \\ \hline
$\mathrm{P1}$ & 0.6097 & 0.9074 & 0.7293 \\ \hline
$\mathrm{P2}$ & 0.2471 & 0.9317 & 0.3906 \\ \hline
$\mathrm{Normal}$ & \cellcolor{gray!15} 0.9985 & 0.9525 & \cellcolor{gray!15} 0.9750 \\ \bottomrule
$\text{weighted avg}$ & \cellcolor{gray!15} 0.9834 & \cellcolor{gray!15} 0.9512 & \cellcolor{gray!15} 0.9636 \\ 
\bottomrule
\end{tabular}
\end{center}
\end{table}

\section{Conclusion}

This study described a methodology to assess the condition of the railway track through degrees of criticality, using acceleration data, vibration modes, and wheel-rail force parameters. The classification model was built based on data generated by multibody simulation, indirectly receiving input from real track geometry data. The simulation and parameters for database generation are continually validated throughout railway concessionaire activities, ensuring reliable data, considering the constraints of the proposed problem, such as the type of wagon, speeds, and analyzed yards.

The results demonstrated a consistent relationship between measured acceleration information and the track condition. The model demonstrates overall high performance, as indicated by its performance across metrics such as sensitivity, precision, and F1-Score, except for the $\mathrm{P2}$ criticality class. The model could classify criticalities, exhibiting lower performance in the $\mathrm{P2}$ class. The results suggest that the limits used in labeling the $\mathrm{P2}$ class could be improved, or the problem could be constrained by removing the $\mathrm{P2}$ class to achieve better overall model performance.

In the context of the given problem, the Logistic-NARX Multinomial Model Approach was crucial and aided the model in comprehending the dataset. It helped in identifying irrelevant features for the model while pinpointing those that contribute most significantly to predictive power. The resulting dataset exhibits high potential for improving algorithm performance, reducing processing time, and preventing overfitting to the data sample. In practical terms, this analysis is indispensable for determining which sensors and locations bear the most significance in the analysis of railway dynamics.

\section*{Acknowledgements}
We thank CAPES, CNPq, INERGE, FAPEMIG and
Federal University of Juiz de Fora (UFJF) for the support.

\bibliography{references}             
                                                   







\end{document}